\documentclass[aps,prb,twocolumn,floatfix,citeautoscript,superscriptaddress,longbibliography]{revtex4-1}

\usepackage{amsmath}
\usepackage{amssymb}
\usepackage{amsfonts}
\usepackage{graphicx,graphics}
\usepackage{bm}
\usepackage{multirow}

\newcommand{\ket}[1]{\lvert #1\rangle}
\newcommand{\bra}[1]{\langle#1 \rvert}
\newcommand{\abs}[1]{\lvert #1 \rvert}
\newcommand{\expect}[1]{\langle #1\rangle}
\newcommand{\braket}[2]{\langle #1 \rvert #2\rangle}

\newcommand{\br}{\mathbf{r}}
\newcommand{\bk}{\mathbf{k}}

\begin{document}

\title{Electron and hole transport in disordered monolayer MoS$_2$:
  atomic vacancy-induced short-range and Coulomb disorder scattering}

\author{Kristen Kaasbjerg}
\email{kkaa@dtu.dk}
\affiliation{Center for Nanostructured Graphene (CNG), Department of Physics, 
  Technical University of Denmark, DK-2800 Kongens Lyngby,
  Denmark}
\author{Tony Low}
\affiliation{Department of Electrical and Computer Engineering,
  University of Minnesota, Minneapolis, MN 55455, USA}
\author{Antti-Pekka Jauho}
\affiliation{Center for Nanostructured Graphene (CNG), Department of Physics, 
  Technical University of Denmark, DK-2800 Kongens Lyngby,
  Denmark}
\date{\today}

\begin{abstract}
  Atomic disorder is a common limiting factor for the low-temperature mobility
  in monolayer transition-metal dichalcogenides (TMDs; $MX_2$). Here, we study
  the effect of often occurring atomic vacancies on carrier scattering and
  transport in $p$- and $n$-type monolayer MoS$_2$. Due to charge trapping in
  vacancy-induced in-gap states, both \emph{neutral} and \emph{charged}
  vacancies resembling, respectively, short-range and combined short-range and
  long-range Coulomb scatterers, must be considered. Using the $T$-matrix
  formalism, we demonstrate a strong renormalization of the Born description of
  short-range scattering, manifested in a pronounced \emph{reduction} and a
  characteristic energy dependence of the scattering rate. As a consequence,
  carrier scattering in TMDs with charged vacancies is dominated by the
  long-range Coulomb-disorder scattering, giving rise to a strong
  screening-induced temperature and density dependence of the low-temperature
  carrier mobility. For TMDs with neutral vacancies, the absence of intrinsic
  Coulomb disorder results in significantly higher mobilities as well as an
  unusual density dependence of the mobility which \emph{decreases} with the
  carrier density. Our work illuminates the transport-limiting effects of
  atomic-vacancy scattering relevant for high-mobility TMD devices.
\end{abstract}
\maketitle

\section{Introduction}

Two-dimensional (2D) monolayers of transition metal dichalcogenides (TMDs;
$MX_2$) hold great promise for future electronics and
optoelectronics~\cite{Heinz:ThinMoS2,Kis:MoS2Transistor,Schuller:Photocarrier,avouris20172d}. In
addition, their spin-valley coupling~\cite{Yao:SpinValley,Heinz:Spin} makes them
potential candidates for spin- and valleytronics applications, which among other
things has sparked interest in TMD-based quantum-dot
qubits~\cite{Burkard:TMDs,Yao:Intervalley,Yao:SpinvalleyQubit,Bednarek:Valley,Ensslin:Gate}. For
such purposes, high-mobility samples with long spin and valley life times are
essential.

Like in conventional 2D semiconductor heterostructure systems, disorder sets the
ultimate limit for the achievable low-temperature mobility in monolayer
TMDs~\cite{Hone:Disorder}, most often limited by short-range and Coulomb
disorder
scattering~\cite{Kis:Engineering,Herrero:Intrinsic,Avouris:Electronic,Wang:Towards,Eda:Transport,Eda:Charge,Hone:Multi,Eda:Quantum,Hone:Low}. Only
recently have mobilities exceeding $\sim$ 1000~cm$^{-2}\,$V$^{-1}\,$s$^{-1}$
been
achieved~\cite{Tutuc:Shubnikov,Ensslin:Gate,Dean:Ambi,Tutuc:Large,Ensslin:Interactions}.

On the theoretical side, studies of the transport properties have focused on
semiclassical transport including
electron-phonon~\cite{Kaasbjerg:MoS2,Kaasbjerg:MoS2Acoustic,Dery:SymmetryBased,Kim:Intrinsic}
and charged impurity scattering~\cite{Fischetti:Mobility,Jena:ChargeScattering},
microscopic descriptions of atomic point defects within the Kubo
formalism~\cite{Heine:Defect,Guinea:Effect,Sousa:Anomalous}, and
quantum-transport
studies~\cite{Shen:Intervalley,Xiao:Spin,Guinea:Quantum,Falko:SpinValley,Peeters:Quantum,Burkard:Landau,Sousa:Anomalous,Houzet:Weak}. However,
the impact of atomic disorder on the transport properties is still poorly
understood.

\begin{figure}[!b]
  \centering
  \hspace{1mm}
  \includegraphics[width=0.99\linewidth]{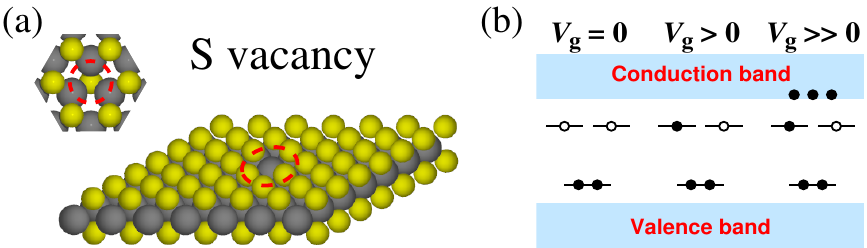}
  \caption{Atomic vacancies in 2D TMDs. (a) Top and side views of the atomic
    structure of a sulfur vacancy in 2D MoS$_2$. The $C_{3v}$ symmetry of the
    vacancy site results in a protection against intervalley
    scattering~\cite{Jauho:Symmetry}. (b) Sulfur vacancy-induced in-gap
    states. In the undoped material ($V_\text{g}=0$), the state close to the
    valence band maximum is occupied with two electrons from the missing sulfur
    anion. With an applied gate voltage, $V_\text{g}>0$, one electron is trapped
    in the upper in-gap states before carriers are introduced into the
    conduction band ($V_\text{g}\gg 0$).}
\label{fig:Svacancy}
\end{figure}
In experimental STM studies on monolayer TMDs, atomic monovacancies have been
found to be among the dominating sources of intrinsic lattice
disorder~\cite{Idrobo:Intrinsic,Suenaga:Threefold,Zhang:Exploring,Wu:Defect,Pasupathy:Approaching,Bargioni:Large}.
Their stability and electronic structure have been studied in great detail
theoretically~\cite{Krasheninnikov:Two,Robertson:Sulfur,Kim:Stability,Nieminen:Charged,Neto:Donor,Krasheninnikov:Native,Sanyal:Systematic,Thygesen:Defect,Leuenberger:Electronic},
demonstrating that they often introduce in-gap states, as illustrated in
Fig.~\ref{fig:Svacancy} for a S vacancy in MoS$_2$, which are localized at the
vacancy site. In-gap states play a crucial role in the transport properties of
2D TMDs as they can trap charges. At low gate-induced doping levels, electrons
become trapped in empty in-gap states and variable-range hopping between the
defect sites dominates the
transport~\cite{Ghosh:Nature,Wang:Hopping,Wang:Towards,Wang:Probing,Morpurgo:Hole}. With
increasing doping, the material becomes $n$-type with free carriers in the
conduction band. At this point, a transition from the insulating state with
thermally-activated hopping transport to a metallic conduction regime with a
conductivity (resistivity) that decreases (increases) with increasing
temperature takes
place~\cite{Ghosh:Nature,Herrero:Intrinsic,Kis:Engineering,Eda:Transport,Wang:Towards,Wang:High,Wang:Realization,Wang:Analyzing}.
Usually, metallic behavior is attributed to electron-phonon scattering, but this
is suppressed at low temperature and disorder scattering by, e.g., atomic
vacancies becomes the mobility-limiting
factor~\cite{Wang:Towards,Eda:Charge,Eda:Transport,Hone:Multi,Eda:Quantum}.

In this work, we provide an in-depth study of the effect of atomic vacancies on
carrier scattering and transport in $p$- and $n$-type 2D
MoS$_2$. Conventionally, atomic vacancies in 2D TMDs are treated as
\emph{neutral} point defects (see, e.g., Ref.~\onlinecite{Guinea:Effect}) which
act as short-range scatterers. However, in the presence of vacancy-induced
in-gap states, vacancies may acquire a dual character due to charging of the
vacancy site, and should be treated as combined short-range and long-range
Coulomb (i.e., charged impurity) scatterers.

The situation for an S vacancy in 2D
MoS$_2$~\cite{Krasheninnikov:Two,Robertson:Sulfur,Kim:Stability,Nieminen:Charged,Neto:Donor,Krasheninnikov:Native,Sanyal:Systematic,Thygesen:Defect,Leuenberger:Electronic}
is illustrated in Fig.~\ref{fig:Svacancy} where the vacancy gives rise to three
in-gap states. The lowest state is doubly occupied in the undoped, charge
neutral material ($V_\text{g}=0$), and hence behaves as a donor state which
traps holes in $p$-doped materials. On the other hand, the upper states are
empty and behave as deep single-electron acceptors which trap electrons in
$n$-doped ($V_\text{g}>0$)
samples~\cite{Robertson:Sulfur,Kim:Stability,Wang:Hopping,Wang:Towards} (doubly
charged vacancy sites are expected to be prohibited by a large onsite Coulomb
repulsion energy). Thus, S vacancies are expected to acquire a net charge in
$p$- or $n$-doped MoS$_2$ (positive and negative, respectively) due to charging
of the in-gap states. The same holds for Mo vacancies in MoS$_2$ (see, e.g.,
Ref.~\onlinecite{Sanyal:Systematic}). However, while vacancies in 2D TMDs, in
general, seem to introduce empty in-gap states, occupied states above the
valence-band edge are not always
present~\cite{Sanyal:Systematic,Huis:Strong}. In the latter case, the vacancy
will remain neutral in the $p$-doped material.

Based on the $T$-matrix formalism~\cite{Rammer,Flensberg}, we here demonstrate
that the scattering amplitude for the short-range component of the scattering
potential is strongly renormalized with respect to its value in the Born
approximation. For charged vacancies, this renders the short-range potential
irrelevant in comparison to the Coulomb contribution to the scattering
potential, and effectively reduces the vacancies to charged impurities.

The dominance of Coulomb disorder scattering gives rise to a strong temperature
and density dependence of the mobility which stems from the temperature
dependent screening of the Coulomb potential. In the degenerate low-temperature
($T\ll T_F$) regime, Coulomb disorder resembles short-range disorder due to the
efficient carrier screening in 2D TMDs, and yields a $\mu\sim n^0,T^0$
behavior. At the crossover to the nondegenerate ($T \gtrsim T_F$) regime, the
screening efficiency is strongly reduced due to the temperature dependence of
the 2D screening function, and the experimentally observed metallic behavior
with a mobility that decreases with increasing temperature and increases with
the carrier density prevails. While this behavior is inherent to any 2D
semiconductor systems at the quantum-classical crossover ($T=T_F$) between the
degenerate and nondegenerate regimes~\cite{Hwang:Screening}, the temperature and
density interval where the crossover takes place is highly dependent on material
parameters (spin and valley degeneracy, effective mass). For typical carrier
densities ($10^{11}$--$10^{13}$~cm$^{-2}$) in 2D TMDs, the Fermi temperature
falls in the range $1\,\mathrm{K} \lesssim T_F \lesssim 100\,\mathrm{K}$, thus
placing the quantum-classical crossover in an easily accessible temperature
range where phonon scattering is
weak~\cite{Kaasbjerg:MoS2,Kaasbjerg:MoS2Acoustic}.

In 2D TMDs where the vacancies remain \emph{neutral} in the $p$-doped material,
the mobility is demonstrated to be significantly higher and shows much weaker
temperature dependence as well as a qualitatively different density dependence
with a mobility that \emph{decreases} with the carrier density. This behavior is
inherent to atomic-vacancy limited transport in 2D TMDs with charge-neutral
vacancies.

Our findings are of high relevance for the understanding of the microscopic
factors governing the mobility, magneto and quantum transport in 2D
TMDs~\cite{Kis:Electrical,Hone:Multi,Eda:Quantum,McEuen:The,Tutuc:Shubnikov,Hone:Low,Ensslin:Gate,Dean:Ambi,Tutuc:Large,Ensslin:Interactions,Eda:Phase}.

\section{Low-energy Hamiltonian and Boltzmann transport theory}
\label{sec:lowenergy}

\begin{figure}[!t]
  \centering
  \includegraphics[width=0.35\linewidth]{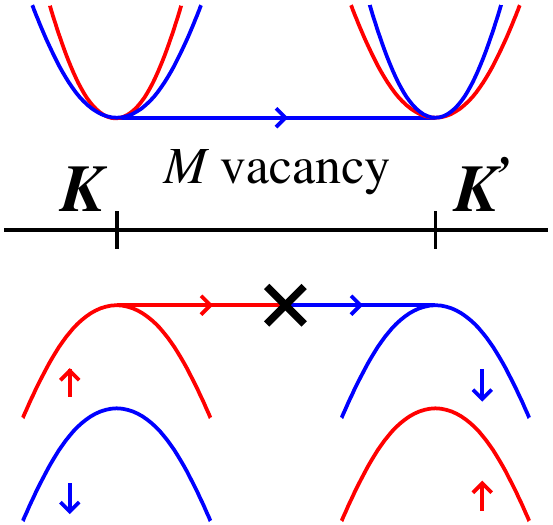}
  \caption{Band structure and disorder scattering in the $K,K'$ valleys of 2D
    TMDs. While intravalley scattering is allowed (not shown), intervalley
    scattering is strongly suppressed by (i) the large spin-orbit splitting in
    the valence band (for some TMDs also in the conduction band), and (ii) a
    symmetry-induced selection rule for defects with $C_3$ symmetry and centered
    at the $M$ and $X$ sites in both the valence and conduction
    bands~\cite{Jauho:Symmetry}.$\,$}
\label{fig:spinvalley}
\end{figure}

In 2D TMDs, the $K,K'$ valence and conduction band valleys illustrated in
Fig.~\ref{fig:spinvalley}, can be described by the effective low-energy
Hamiltonian~\cite{Yao:SpinValley}
\begin{align}
  \label{eq:H}
  \mathcal{H}_\tau(\bk) & = at
     \left( \tau k_x \hat{\sigma}_x + k_y \hat{\sigma}_y \right)
     + \frac{\Delta}{2} \hat{\sigma}_z  
     + \tau \frac{\Delta_\text{SO}}{2}\frac{\hat{\sigma}_0 - \hat{\sigma}_z}{2}
                          \otimes \hat{s}_z  ,
\end{align}
where $a$ is the lattice parameter, $t$ a hopping parameter, $\tau=\pm 1$ is the
$K,K'$ valley index, $\Delta$ is the band gap, $\Delta_\text{SO}$ ($=148$~meV in
2D MoS$_2$) is the spin-orbit (SO) induced spin splitting at the top of the
valence band~\cite{Schwing:GiantSO}, and $\hat{\sigma}_i$, $\hat{\tau}_i$ and
$\hat{s}_i$, are the identity ($i=0$) and Pauli matrices $(i=x,y,z$) in the
symmetry-adapted orbital basis, valley and spin, respectively. The orbital basis
$\{\ket{\phi_{\sigma\tau}}\}$ is spanned by the $M$ $d$-orbitals
$\ket{\phi_{v\tau}} = 1/\sqrt{2} \left( \ket{d_{x^2 - y^2}} + i \tau\ket{d_{xy}}
\right)$
and $\ket{\phi_{c\tau}}=\ket{d_{z^2}}$. 

In the continuum description of Eq.~\eqref{eq:H}, the wave function can be
expressed as (suppressing the spin index $s=\pm 1$ for brevity)
$\psi_{\sigma\tau\bk}(\br) = \tfrac{1}{\sqrt{A}}e^{i\bk\cdot \br}
\chi_{\sigma\tau}$
where $A$ is the sample area and the valence ($\sigma=v$) and conduction
($\sigma=c$) band eigenspinors $\chi_{\sigma\tau}$ are dominated by,
respectively, $\ket{\phi_{v\tau}}$ and $\ket{\phi_{c\tau}}$ near the band
edges~\cite{Xiao:Spin}. In this energy range, the bands are parabolic
$\varepsilon_{\sigma\tau\bk} = E_{\sigma\tau} \pm \hbar^2 k^2 / 2 m_\sigma^*$,
where $E_{\sigma\tau} = \pm \Delta/2 + \delta_{v\sigma} \tau \Delta_\text{SO}/2$
is the band-edge position (the spin dependence is shown in
Fig.~\ref{fig:spinvalley}). The small SO splitting of 2--3~meV in the conduction
band of MoS$_2$~\cite{Falko:Monolayer,Rossier:Large} is here neglected and spin
degeneracy is assumed. For the effective masses, we use our calculated DFT
values, $m_v^*=0.59$ and $m_c^* = 0.48$ for 2D MoS$_2$~\cite{calculations},
which\nocite{GPAW,GPAW1,GPAW2} are in good agreement with previously reported
values~\cite{Lambrecht:Quasiparticle,Kaasbjerg:MoS2,Falko:Monolayer}.

\subsection{Disorder-limited transport}

In Boltzmann transport theory, the carrier mobility
$\mu_{xx} = \sigma_{xx} / ne$, where $\sigma_{xx}$ is the conductivity, limited
by elastic disorder scattering can be expressed as
$\mu_{xx} = e \expect{\tau_k}/m^*$, where $m^*$ is the effective mass and
$\tau_k$ is the total energy-dependent momentum relaxation time given by the sum
over contributions from different types of disorder,
$\tau^{-1} = \sum_i \tau_i^{-1}$. The brackets $\expect{\cdot}$ denote the
energy-weighted average defined as
\begin{equation} 
  \label{eq:tau_averaged}
  \expect{A} = \frac{1}{n}
  \int \! d\varepsilon \; \rho_\sigma(\varepsilon) 
      \varepsilon A(\varepsilon)
 \left(- \frac{\partial f}{\partial \varepsilon} \right).
\end{equation}
Here, $n$ is the two-dimensional carrier density,
$\rho_\sigma = g_s g_\tau m_\sigma^* / 2\pi\hbar^2$ is the constant density of
states in 2D, $g_s$ and $g_\tau$ are the spin ($g_s=1,2$ for $\sigma=v,c$) and
valley degeneracy, respectively,
$f(\varepsilon) = \left\{1 + \exp{[ (\varepsilon - \mu)/ k_\text{B} T ]}
\right\}^{-1}$
is the equilibrium Fermi-Dirac distribution function, and $\mu$ is the chemical
potential. For a degenerate electron gas, scattering is restricted to a thin
shell of width $k_\text{B} T$ around the Fermi level $E_F$ and
$\mu_{xx} \approx e \tau(E_F) / m^*$.

For random disorder, the relaxation time due to intra ($\tau=\tau'$) and
intervalley ($\tau\neq\tau'$) scattering off defects of type $i$ is in the Born
approximation given by
\begin{align}
  \label{eq:tau}
  \frac{1}{\tau_{i,\sigma\tau}(\varepsilon_\bk)} 
  & = \frac{2\pi}{\hbar} n_i  \sum_{\tau'}
       \int \! \frac{d\bk'}{(2\pi)^2} \,
           \abs{V_{\bk\bk'}^{i,\sigma}(\tau,\tau')}^2
       \nonumber \\ & \quad \times 
       \left( 1 - \cos \theta_{\bk\bk'} \right) 
       \delta(\varepsilon_\bk - \varepsilon_{\bk'})  ,
\end{align}
where $n_i$ is the disorder density,
$V_{\bk\bk'}^{i,\sigma}(\tau,\tau')=
\bra{\chi_{\sigma\tau}}\hat{V}_i\ket{\chi_{\sigma\tau'}}$
is the matrix element (in units of $\mathrm{eV}\,\mathrm{\AA}^2$) of the defect
scattering potential with respect to the electronic states in band $\sigma$ and
with wave vectors $\bk,\bk'$, and $\theta_{\bk\bk'}$ is the scattering angle.
% In the continuum description, the matrix elements can be expressed as
% \begin{equation}
%   \label{eq:M}
%   V_{\bk\bk'}^{i,\sigma}(\tau,\tau') 
%   % = \bra{\psi_{\sigma\tau\bk}} \hat{V}_i \ket{\psi_{\sigma\tau'\bk'}}
%   % = V_\bq^i \braket{\chi_{\sigma\tau}}\hat{V}_i{\chi_{\sigma\tau'}}
%   = V_\bq^i \bra{\chi_{\sigma\tau}}\hat{V}_i\ket{\chi_{\sigma\tau'}}
% \end{equation}
% where $V_\bq^i$ is the 2D Fourier transform of the defect potential,
% $\bq=\bk'-\bk + (\tau'-\tau) \bK$, and we have taken the idealized
% zero-thickness 2D limit which is reasonable for atomically thin 2D materials. 

Here, we omit spin-flip scattering and limit the discussion to spin-conserving
scattering off nonmagnetic defects, $\hat{V}_i\propto \hat{s}_0$. Due to the
orthogonality of the $K,K'$ valence-band orbitals,
$\braket{\phi_{v\tau}}{\phi_{v\tau'}} = \delta_{\tau\tau'}$, the valence-band
intervalley matrix element is in general suppressed~\cite{Xiao:Spin}. Recently,
we have demonstrated a stricter symmetry-induced selection rule for defects with
$C_3$ symmetry which applies to the intervalley matrix element in both the
valence and conduction bands~\cite{Jauho:Symmetry} (see also
Sec.~\ref{sec:tmatrix} below). As a consequence, intervalley scattering in 2D
TMDs is only allowed in the conduction band for metal-centered defects as
illustrated in Fig.~\ref{fig:spinvalley}.

Conventionally, point defects such as atomic vacancies are treated as
short-range scatterers described by a $\delta$-function scattering potential in
real space, $V_i(\br) = V_{0,\sigma}^i\delta(\br)$ where $V_{0,\sigma}^i$ is a
band-dependent disorder strength. In this case, the matrix element is simply
given by
\begin{equation}
  \label{eq:V0}
  V_{\bk\bk'}^{i,\sigma}(\tau,\tau') 
  = V_{0,\sigma}^i (\tau,\tau') ,
\end{equation}
where the valley dependence accounts for the above-mentioned selection rules.
In the Born approximation, this yields a momentum relaxation time,
$\tau_i^{-1} = n_i m^* \abs{V_{0,\sigma}^i}^2 /\hbar^3$, and mobility,
$\mu\sim n^0$, which are independent on the carrier energy and carrier density,
respectively. For atomic vacancies in 2D TMDs this approach may break down for two
reasons. 

First of all, as discussed in the introduction, charging of the vacancy sites in
the presence of in-gap states gives rise to two distinct contributions to
the scattering potential given, respectively, by (i) a short-range potential
($V_0^i$) associated with the atomic-scale point defect created by the vacancy,
and (ii) a long-range Coulomb potential ($V_C^i$) due to the charged vacancy
site. Here, we assume that the relaxation time for charged vacancies can be
obtained as
\begin{equation}
  \label{eq:tau_total}
  \tau_{i}^{-1} = \tau_{i,0}^{-1} + \tau_{i,C}^{-1} ,
\end{equation}
where the two relaxation times on the right-hand side account for the
above-mentioned contributions to the scattering potential.

Secondly, the Born approximation only applies to \emph{weak} short-range
disorder, i.e. when $V_0^i$ is small (see discussion in Sec.~\ref{sec:tmatrix}),
and may break down for vacancies. For disorder of arbitrary strength, the
\emph{bare} matrix element in Eq.~\eqref{eq:tau} must be replaced with the $T$
matrix, i.e.
$V_{\bk\bk'}^i \rightarrow T_{\bk\bk'}^i(\varepsilon_\bk)$~\cite{Flensberg}. The
$T$ matrix solves the single-defect problem exactly by taking into account
multiple scattering off the individual defects, and is hence \emph{exact} for
dilute disorder, i.e. $n_i\ll 1/A_\text{cell}$ where $A_\text{cell}$ is the
unit-cell area.

In the following subsections, we analyse the scattering properties of atomic
vacancies (Mo and S) in 2D TMDs.

% \section{Atomic vacancies: midgap states and scattering properties}
% \label{sec:TMDs}
% 
% As representatives of the most common types of atomic defects in 2D TMDs, we
% consider in this work Mo and S monovacancies in 2D MoS$_2$. As we have pointed
% out in the introduction, empty (occupied) vacancy-induced midgap states become
% filled (emptied) when electrons (holes) are introduced in the conduction
% (valence) band. This gives rise to two distinct contributions to the vacancy
% scattering potential given, respectively, by (i) a short-range potential
% ($V_0^i$) associated with the atomic-scale point defect created by the vacancy,
% and (ii) a long-range Coulomb potential ($V_C^i$) due to the charged vacancy
% site. As a result, atomic vacancies in 2D TMDs act as combined short-range and
% Coulomb disorder scattering centers.
% 
% We assume that the relaxation time for charged vacancies can be obtained as
% \begin{equation}
%   \label{eq:tau_total}
%   \tau_{i}^{-1} = \tau_{i,0}^{-1} + \tau_{i,C}^{-1} ,
% \end{equation}
% where the relaxation times on the right-hand side are due to the two
% above-mentioned contributions to the scattering potential. Their forms and
% scattering properties are outlined in the following two subsections.

\subsection{Short-range potential: $V_0^i$}
\label{sec:V0}

Short-range disorder due to atomic point defects often act as strong scattering
centers, and must be treated with the $T$-matrix formalism which describes
multiple scattering off the same defect to infinite order in the scattering
potential~\cite{Flensberg}. Here, we combine atomistic calculations with an
analytic approach to get a realistic description of the $T$ matrix scattering
rate.

\subsubsection{Low-energy $T$-matrix model}
\label{sec:tmatrix}

For a general defect characterized by the matrix element
$V_{\bk\bk'}^{\sigma, i}$ of its scattering potential between the Bloch states
with wave vectors $\bk,\bk'\in 1$st Brillouin zone (BZ), the $T$ matrix (for a
given spin) is given by the integral equation
\begin{equation}
  \label{eq:Tmatrix}
  T_{\bk\bk'}^{\sigma, i} (\varepsilon) 
  = V_{\bk\bk'}^{\sigma, i}
      + \int_{_{\text{BZ}}} \! \frac{d\bk_1}{(2\pi)^2} 
      V_{\bk\bk_1}^{\sigma, i}
      G_{\sigma \bk_1 }^0(\varepsilon) 
      T_{\bk_1\bk'}^{\sigma, i} (\varepsilon) ,
\end{equation}
where
$G_{\sigma \bk}^0 (\varepsilon) = (\varepsilon - \varepsilon_{\sigma \bk} +
i\eta)^{-1}$
is the \emph{bare} Green's function and the integral is over the 1st~BZ.

With the bandstructure described by the low-energy model in Eq.~\eqref{eq:H}, we
confine the $\bk_1$ integral in~\eqref{eq:Tmatrix} to the $K,K'$ valleys, i.e.
$\int_{_{\text{BZ}}} \! \tfrac{d\bk}{(2\pi)^2} \rightarrow \sum_\tau \int_{\tau}
\!  \tfrac{d\bk}{(2\pi)^2}$,
and parametrize the defect matrix element by a band- and valley-dependent
disorder strength $V_{0,\sigma}^i$ like in Eq.~\eqref{eq:V0}. With this, the
$T$-matrix equation in~\eqref{eq:Tmatrix} reduces to a $\bk,\bk'$-independent
algebraic $2\times2$ matrix equation in the valley indices,
\begin{equation}
  \label{eq:T}
  \hat{T}_\sigma^i (\varepsilon) 
  = \left[ \hat{\tau}_0 - \hat{\bar{G}}_\sigma(\varepsilon) \hat{V}_{0,\sigma}^i
     \right]^{-1} \hat{V}_{0,\sigma}^i ,
\end{equation}
where
$\hat{\bar{G}}_{\sigma}(\varepsilon)=\mathrm{diag} \left[ \bar{G}_{\sigma
  K}(\varepsilon),\bar{G}_{\sigma K'}(\varepsilon) \right]$
is diagonal with elements given by the $\bk$-integrated valley GF,
$\bar{G}_{\sigma\tau}(\varepsilon) = \int_\tau \! \tfrac{d\bk}{(2\pi)^2}
G_{\sigma\tau\bk}^0(\varepsilon)$,
and the diagonal (off-diagonal) elements of $\hat{V}_{0,\sigma}^i$ correspond to
intravalley (intervalley) couplings.

In a recent work~\cite{Jauho:Symmetry}, we have shown that the
$K\leftrightarrow K'$ intervalley coupling due to defects in 2D TMDs with $C_3$
symmetry is suppressed by a symmetry-induced selection rule, except for
$M$-centered defects where intervalley coupling is possible in the conduction
band as illustrated in Fig.~\ref{fig:spinvalley}. As the intra- and intervalley
matrix elements for $M$ vacancies are comparable~\cite{Jauho:Symmetry}, a single
disorder strength can therefore be considered for both $M$ and $X$
vacancies. The defect potential can hence be written
\begin{align}
  \label{eq:Vmatrix}
  \hat{V}_{0,\sigma}^i  = V_{0,\sigma}^i \times
  \left\{
    \begin{array}{l}
      \hat{\tau}_0 + \hat{\tau}_x
      , \quad \sigma = c,\, i=M\\
      \quad \\
      \hat{\tau}_0
      , \quad \text{otherwise} ,
    \end{array}
  \right. 
\end{align}
which captures the defect and band dependent suppression of the intervalley
coupling described above.

With the form of the defect potential in~\eqref{eq:Vmatrix}, the $T$ matrix 
in~\eqref{eq:T} simplifies considerably. For the two cases
in~\eqref{eq:Vmatrix}, we get
\begin{align}
  \label{eq:Tmatrix_2x2}
  \hat{T}_\sigma^i (\varepsilon) = 
  \left\{
    \begin{array}{l}
      T_{c K}^i(\varepsilon) 
      (\hat{\tau}_0 + \hat{\tau}_x) \\
      % , \quad \sigma = c,\, i=M\\
      \quad \\
      T_{\sigma K}^i(\varepsilon) \frac{(\hat{\tau}_0 + \hat{\tau}_z)}{2} +
      T_{\sigma K'}^i(\varepsilon) \frac{(\hat{\tau}_0 - \hat{\tau}_z)}{2}
      , %\, \text{otherwise} ,
    \end{array}
  \right. 
\end{align}
where
\begin{equation}
  \label{eq:Tmatrix_V0}
  T_{\sigma\tau}^i(\varepsilon) = \frac{V_{0,\sigma}^i}
      {1 - g_{\sigma i} V_{0,\sigma}^i \bar{G}_{\sigma\tau}(\varepsilon)} ,
\end{equation}
is the band and valley dependent $T$ matrix with $g_{\sigma i}$ denoting a
valley multiplication factor given by the number of inequivalent valleys coupled
by the defect type $i$, i.e. $g_{c M}=2$ and $g_{\sigma i}=1$ otherwise. For
the parabolic bands in the $K,K'$ valleys, the $\bk$-integrated Green's function
$\bar{G}_{\sigma\tau}$ for $\sigma = c,v$ becomes
\begin{align}
  \label{eq:Gbar}
  \bar{G}_{\sigma\tau}(\varepsilon) & = \pm \bar{\rho}_\sigma
    \ln\frac{\abs{\varepsilon - E_{\sigma\tau}}}
            {\abs{\varepsilon - E_{\sigma\tau} \mp \Lambda}} 
    -i \pi \bar{\rho}_\sigma \theta(\varepsilon \mp E_{\sigma\tau}),
\end{align}
where $\bar{\rho}_\sigma = m_\sigma^*/2\pi\hbar^2$ is the density of states
excluding spin and valley degeneracy, $E_{\sigma\tau}$ is the valley (and spin)
dependent band extrema, and $\Lambda$ is an ultraviolet cut-off to be determined
below. Note that due to the neglected spin-orbit splitting in the conduction band, i.e.
$E_{cK}=E_{cK'}$, the elements of the $T$ matrix in the upper equation in
Eq.~\eqref{eq:Tmatrix_2x2} become independent on the valley index.

Via the optical theorem~\cite{Rammer,Flensberg}, $T_{\sigma\tau}^i(\varepsilon)$
in Eq.~\eqref{eq:Tmatrix_V0} can be identified as the scattering amplitude in
the $T$-matrix approximation. Together with Eq.~\eqref{eq:Gbar}, the formal
condition for disorder to be weak (strong) thus becomes
$V_{0,\sigma}^i \bar{\rho}_\sigma \ll 1$ ($\gg 1$). In monolayer MoS$_2$
$\bar{\rho}_\sigma\approx m^*_\sigma \times 0.02 $~eV$^{-1}$\AA$^{-2}$, implying
that $V_{0,\sigma}^i \gg / \ll 100$~eV$\,$\AA$^2$ corresponds to strong/weak
disorder. For weak disorder, the Born approximation for the scattering
amplitude, $T_{\sigma}^i(\varepsilon)\approx V_{0,\sigma}^i$, is recovered,
whereas the $T$ matrix becomes independent on the disorder potential,
$T_\sigma^i(\varepsilon)\approx -1/\bar{G}_\sigma(\varepsilon)$, for strong
disorder in the \emph{unitary} limit, $V_0^i\rightarrow \infty$. In the
\emph{intermediate} regime, the $T$ matrix should also be considered as it gives
rise to a nonnegligible renormalization of the Born scattering amplitude.

The $T$-matrix renormalization of the Born scattering amplitude manifests itself
in an inherent energy dependence of the scattering rate. For the $T$-matrix in
Eq.~\eqref{eq:Tmatrix_V0}, the inverse momentum relaxation time in
Eq.~\eqref{eq:tau} becomes,
\begin{equation}
  \label{eq:tau_tmatrix}
  \frac{1}{\tau_{i,\sigma\tau} (\varepsilon)} = \frac{2\pi}{\hbar} 
  \frac{n_i  g_{\sigma i}\bar{\rho}_\sigma \abs{V_{0,\sigma}^i}^2}
       {\left(1 \mp g_{\sigma i} \bar{\rho}_\sigma V_{0,\sigma}^i 
           \ln \frac{\abs{\varepsilon-E_{\sigma\tau}}}{\Lambda}\right)^2
         + \left(\pi g_{\sigma i} \bar{\rho}_\sigma V_{0,\sigma}^i \right)^2},
\end{equation}
% where the energy dependence in the denominator of the logarithm in
% $\bar{G}_{\sigma\tau}$ in Eq.~\eqref{eq:Gbar} has been omitted. 
which for the constant defect potentials considered here, is identical to the
quantum scattering rate related to the imaginary part of
the $T$ matrix [see Eq.~\eqref{eq:gamma} below], and given by Eq.~\eqref{eq:tau}
with the replacement $( 1 - \cos \theta_{\bk\bk'} ) \rightarrow 1$. Thus
$\tau_{i,\sigma\tau}^{-1} (\varepsilon) = - \tfrac{2}{\hbar} n_i \mathrm{Im}
T_{\sigma\tau}^i (\varepsilon) = \tfrac{2 \pi}{\hbar} n_i g_{\sigma i}
\bar{\rho}_\sigma \abs{T_{\sigma\tau}^i(\varepsilon)}^2$
where the factor $g_{\sigma i} \bar{\rho}_\sigma$ stems from the $\bk'$ integral
in Eq.~\eqref{eq:tau}.

Figure~\ref{fig:Tmatrix} shows the energy dependence of the inverse relaxation
time in Eq.~\eqref{eq:tau_tmatrix} (full lines) in terms of
$-\mathrm{Im}T_{\sigma\tau}^i(\varepsilon)$ for parameters corresponding to Mo
and S vacancies in MoS$_2$ (see Sec.~\ref{sec:fit} below). The characteristic
behavior of the scattering rate which increases away from the band edge where it
drops sharply, can be traced back to $\bar{G}_\sigma$ in~\eqref{eq:Gbar} which
diverges for $\abs{\varepsilon - E_\sigma} \rightarrow 0$ due to the logarithm
in its real part. This behavior is inherent to \emph{intermediate} and
\emph{strong} short-range disorder, such as, e.g., vacancies and point defects,
in 2D materials with parabolic bands, and in stark contrast to the constant
relaxation time in the Born approximation,
$\tau_{i,\sigma\tau}^{-1} \approx \tfrac{2\pi}{\hbar} n_i g_{\sigma i}
\bar{\rho}_\sigma \abs{V_{0,\sigma}^i}^2$,
which is recovered for \emph{weak} disorder.

Also, the above is in contrast to the situation for vacancies in graphene where
the scattering rate is predicted to have a nonmonotonic energy dependence due to
resonance scattering off quasibound defect states in the vicinity of the Dirac
point~\cite{Mirlin:Electron,Guinea:Electronic,Basko:Resonant,Katsnelson:Resonant,Kaasbjerg:First}.

\subsubsection{Effective disorder parameters for vacancies}
\label{sec:fit}

To determine the $T$-matrix parameters for Mo and S vacancies which enter
Eq.~\eqref{eq:Tmatrix_V0}, we have used the atomistic method outlined in
Refs.~\onlinecite{Jauho:Symmetry,Kaasbjerg:First} to calculate the $T$ matrix
for atomic vacancies in 2D MoS$_2$ based on Eq.~\eqref{eq:Tmatrix} with DFT
inputs for the band structure (including spin-orbit interaction) and defect
matrix elements sampled in the entire BZ~\cite{calculations}.

In order to quantify our atomistic calculations, we consider the quasiparticle
scattering rate $\tau_{i,\sigma\bk}^{-1} = \gamma_{\sigma\bk}^i / \hbar$, where
\begin{equation}
  \label{eq:gamma}
  \gamma_{\sigma\bk}^i = -2 \mathrm{Im}\, \Sigma_{\sigma\bk}^i(\varepsilon_{\sigma\bk})
                = - 2 n_i \mathrm{Im}\, T_{\bk\bk}^{\sigma,i}(\varepsilon_{\sigma\bk}) ,
\end{equation}
is the lifetime broadening given by the imaginary part of the on-shell
($\varepsilon=\varepsilon_{\sigma \bk}$) $T$-matrix self-energy
$\Sigma_{\sigma\bk}^i = n_i T_{\bk\bk}^{\sigma,i}$ of the disorder-averaged
Green's function~\cite{Rammer,Flensberg}.
\begin{figure}[!t]
  \centering
  \includegraphics[width=0.99\linewidth]{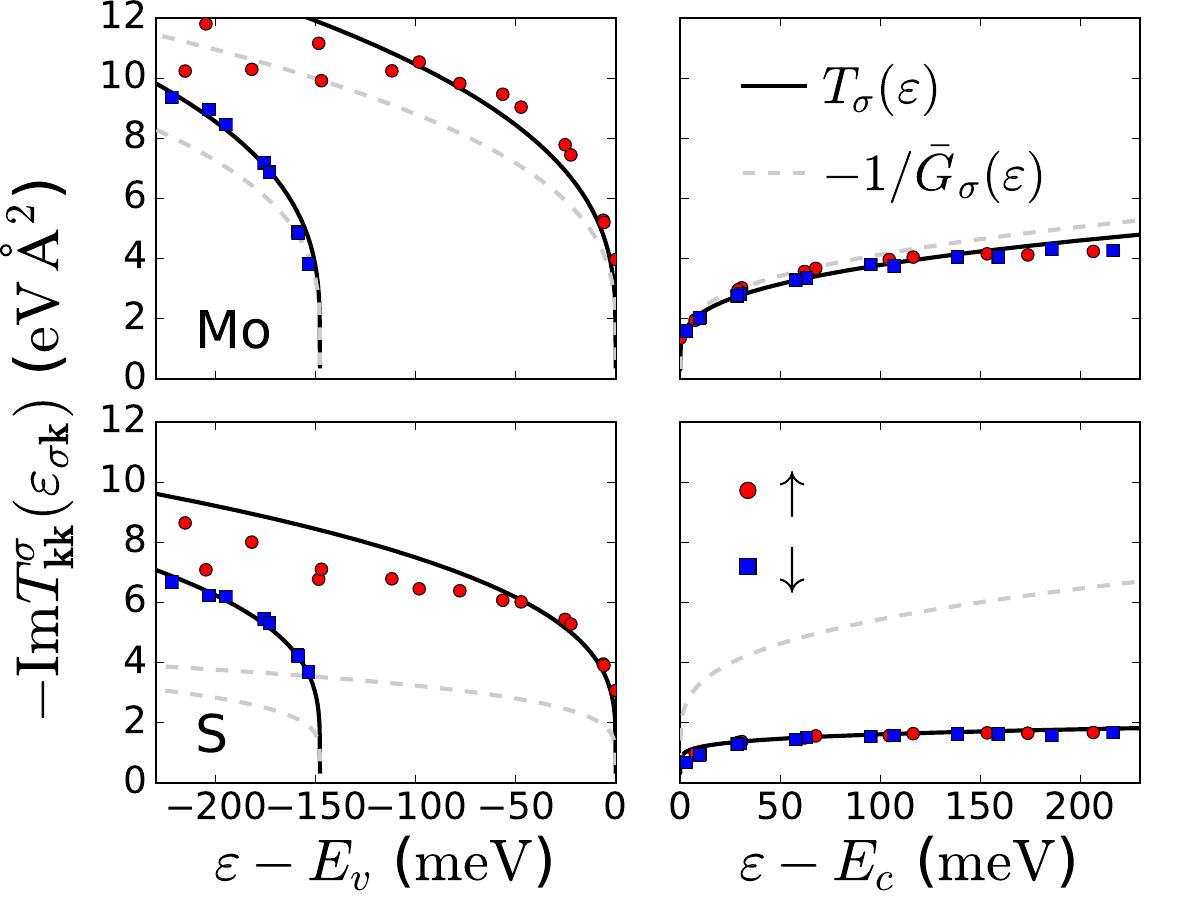}
  \caption{Imaginary part of the on-shell $T$ matrix as a function of energy in
    the $K$ valley for the valence ($\sigma=v$, left) and conduction
    ($\sigma=c$, right) band, and Mo (top) and S (bottom) vacancies. The energy
    is measured with respect to the band edges $E_\sigma$. The symbols show the
    $T$ matrix obtained with the atomistic DFT method in
    Ref.~\onlinecite{Kaasbjerg:First} (along the path $\Gamma$-$K$-$M$ in the BZ
    sampled with $135\times 135$ $\bk$ points). The solid (dashed) lines show
    the analytic $T$ matrix in Eq.~\eqref{eq:Tmatrix_V0} (unitary limit) with
    the parameters fitted to the atomistic results as described in
    Sec.~\ref{sec:fit} (see also Tab.~\ref{tab:parameters}). Due to the large
    spin-orbit splitting in valence band, only spin-up points appear at low
    energies $\abs{\varepsilon - E_v} < \Delta_\text{SO}$ where
    $\Delta_\text{SO}= 148$~meV in 2D MoS$_2$.~\cite{Schwing:GiantSO} }
\label{fig:Tmatrix}
\end{figure}

Figure~\ref{fig:Tmatrix} shows the imaginary part of the calculated on-shell $T$
matrix for Mo (top) and S (bottom)
vacancies~\cite{Jauho:Symmetry,Kaasbjerg:First} as a function energy in the $K$
valley of the valence (left) and conduction (right) band. The red dots and blue
squares correspond to the $T$-matrix for the spin-orbit split spin up and down
bands, respectively. In the plots for the valence band, the offset in energy
between the red dots and blue squares is due to the large spin-orbit splitting
between the spin up and down bands ($\sim 148$~meV) in the $K,K'$ valleys. In
the conduction band, the small spin-orbit splitting of $\sim 2$--3~meV is hardly
discernible, and the imaginary part of the $T$ matrix is almost identical for
the spin up and down bands. The $T$-matrix results in Fig.~\ref{fig:Tmatrix} can
be converted to a lifetime broadening via Eq.~\eqref{eq:gamma}, and correspond
to a broadening of the order of $\gamma_{\sigma\bk}^i\sim 0.01$--0.1~meV for a
disorder density of $n_i=10^{11}\,\mathrm{cm}^{-2}$. Furthermore, the atomistic
$T$-matrix results confirm the characteristic energy dependence of the lifetime
broadening, or scattering rate, predicted by Eq.~\eqref{eq:tau_tmatrix}.
\begin{table}[!t]
\begin{ruledtabular}
\begin{tabular}{lcc}
    &  Valence band  & Conduction band  \\ 
\hline                                     
\multirow{ 3}{*}{Mo vacancy}
    &   $V_{0,\sigma}^i=145\,\mathrm{eV}\,\mathrm{\AA}^2$ 
    &   $V_{0,\sigma}^i=155\,\mathrm{eV}\,\mathrm{\AA}^2$   \\
    &   $\Lambda = 8$~eV       
    &   $\Lambda = 20$~eV  \\
    &   $g_\mathrm{\sigma i} = 1$ & $g_{\sigma i} = 2$  \\
\multirow{ 3}{*}{S vacancy} 
    &   $V_{0,\sigma}^i = 24\,\mathrm{eV}\,\mathrm{\AA}^2$ 
    &   $V_{0,\sigma}^i = 15\,\mathrm{eV}\,\mathrm{\AA}^2$  \\
    &   $\Lambda = 400$~eV     
    &   $\Lambda = 100$~eV  \\
    &   $g_{\sigma i} = 1$ &  $g_{\sigma i} = 1$
\end{tabular}
\end{ruledtabular}
\caption{Effective disorder parameters for the analytic $T$-matrix model in
  Eq.~\eqref{eq:Tmatrix_V0} for Mo and S vacancies in MoS$_2$. The
  disorder strengths $V_{0,\sigma}^i$ are fixed to the value of the
  atomistic DFT matrix elements in Ref.~\onlinecite{Kaasbjerg:First}, while the
  energy cutoff is treated as a free parameter in order to fit the 
  atomistic results in Fig.~\ref{fig:Tmatrix} with the analytic model (see
  Sec.~\ref{sec:fit} for details).}
\label{tab:parameters}
\end{table}

The $T$-matrix parameters for vacancies are obtained by fitting the analytic
expression~\eqref{eq:Tmatrix_V0} to the atomistic $T$ matrix in
Fig.~\ref{fig:Tmatrix} [we fit the imaginary parts to achieve an optimal
description of the scattering rate, cf. Eq.~\eqref{eq:gamma}]. In the fitting
procedure, we first fix the disorder strength $V_{0,\sigma}^i$ to the value of
the atomistic matrix elements $V_{\bk\bk'}^{\sigma, i}$ at the $K,K'$ points
(see Refs.~\onlinecite{Jauho:Symmetry,Kaasbjerg:First}), and treat the
ultraviolet cutoff $\Lambda$ as a fitting parameter. This is justified as
$\Lambda$ does not have an immediate physical interpretation in the low-energy
$T$-matrix model. Instead, it should be regarded as an effective parameter which
compensates for approximating in Eq.~\eqref{eq:Tmatrix} (i) the band structure
in the full BZ with parabolic bands in the $K,K'$ valleys, and (ii) the defect matrix
element by a constant disorder strength.

The resulting fits are shown with solid lines (the dashed lines show the unitary
limit) in Fig.~\ref{fig:Tmatrix}, and are seen to match the atomistic
calculations almost perfectly, in particular at energies near the band edges
relevant for transport. The corresponding fitting parameters are summarized in
Table~\ref{tab:parameters}.

As witnessed by Fig.~\ref{fig:Tmatrix}, the $T$ matrix is highly dependent on
the band. This is due to the $\pm 1$ factor in front of the real part of
$\bar{G}$ in Eq.~\eqref{eq:Gbar}, which even for an electron-hole symmetric
bandstructure gives rise to different scattering rates for electrons and holes
[cf. Eq.~\eqref{eq:tau_tmatrix}]. For $V_0^i>0$ (the reverse holds for
$V_0^i<0$), this leads to a hole scattering rate which is larger than the
electron scattering rate as seen in Fig.~\ref{fig:Tmatrix}. For the same reason,
the electron and holes rates are, respectively, larger and smaller than the rate
for unitary scattering.

Finally, we note that the $T$-matrix results in Fig.~\eqref{fig:Tmatrix}
correspond to a significant renormalization of the Born scattering amplitude
given by the bare disorder strengths in Tab.~\ref{tab:parameters}. From the
numbers in Fig.~\ref{fig:Tmatrix}, the $T$-matrix scattering amplitudes are
found to be up to an order of magnitude \emph{smaller} than the Born scattering
amplitudes. The renormalization is most pronounced for Mo vacancies which are
strong (almost unitary) scatterers, and still noticeable for the weaker S
vacancies. As the scattering rate depends on the square of the scattering
amplitude, the Born approximation therefore severely overestimates the scattering
rate.

To summarize, the strong $T$-matrix renormalization of the Born scattering
amplitude results in (i) the characteristic energy dependence of the scattering
rates in Fig.~\ref{fig:Tmatrix}, as well as (ii) a significant reduction of the
scattering rate relative to the Born result. These findings points to a
concomitant breakdown of the Born approximation for atomic-vacancy scattering in
2D TMDs.

\subsection{Long-range Coulomb potential: $V_C^i$}
\label{sec:VC}

The relaxation time in Eq.~\eqref{eq:tau_total} due to trapped charges in the
in-gap states is governed by a long-range Coulomb potential. Due to the
localized nature of the in-gap states, this can be approximated by the screened
Coulomb potential from a point charge located at the vacancy site, and the
matrix element becomes
\begin{equation}
  \label{eq:Mcoul}
  V_C^i(q) = \frac{e^2}{2 \epsilon_0 q \epsilon(q)} .
\end{equation}
where $\epsilon(q)\equiv\epsilon(q,T,\mu)$ is the static dielectric function of
the 2D material which includes both intra and interband screening as discussed
in further detail in Sec.~\ref{sec:dielectric} below.

For scattering off the long-range Coulomb potential, we use the Born
approximation which is justified due to the screening of the Coulomb
potential~\cite{Fischetti:Mobility,Hwang:Universal,Jena:ChargeScattering,Hwang:Short,Hwang:Screening}. For
a degenerate 2DEG, the dielectric function is given by
$\epsilon(q) = \kappa(1 + q_\text{TF}/q)$, $q<2k_F$, where
$q_\text{TF} =\tfrac{g_sg_ve^2m^*}{4\pi\epsilon_0\kappa\hbar^2}$ is the
Thomas-Fermi (TF) wave vector and $\kappa$ is a background dielectric
constant. In 2D TMDs, $q_\text{TF} \sim \pi / a \gg k_F$, where $a$ is the
lattice constant, implying that carrier screening is very efficient. This
changes qualitatively the $1/q$ dependence of the unscreened Coulomb potential,
with the screened Coulomb matrix element in~\eqref{eq:Mcoul} becoming
\begin{equation}
    \label{eq:VC0}
    V_C^i (q) \approx
    \frac{e^2}{2\epsilon_0 q_\text{TF}} =
    \frac{2\pi\hbar^2}{g_sg_v m^*} 
    \equiv V_0^C , \quad q< 2k_F,
\end{equation}
which is independent on $q$, and thus resembles short-range disorder with an
\emph{effective} disorder strength $V_0^C$. For $q>2k_F$, the screening
efficiency is reduced and the unscreened $1/q$ potential is recovered. In 2D
MoS$_2$, the effective disorder strengths in the valence and conduction bands
are $V_{0,v}^C \approx 41~\mathrm{eV}\,$\AA$^2 $ and
$V_{0,c}^C \approx 25~\mathrm{eV}\,$\AA$^2$, respectively.

% Altogether, the use of the Born approximation for Coulomb disorder scattering is
% justified. Nevertheless we expect minor quantitative corrections in a
% $T$-matrix calculation of the scattering rate. Unfortunately the $T$-matrix
% equation~\eqref{eq:Tmatrix} does not lend itself to a simple analytic solution
% for the Coulomb matrix elements in Eq.~\eqref{eq:Mcoul} due to the complex $q$
% dependence introduced by the dielectric function. 

In passing, we note that scattering off an \emph{unscreened} 2D Coulomb potential
was studied with a partial-wave analysis (which is equivalent to a $T$-matrix
treatment) in Ref.~\onlinecite{Lin:Scattering}. However, as it has been shown in
numerous
works~\cite{Fischetti:Mobility,Hwang:Universal,Jena:ChargeScattering,Hwang:Short,Hwang:Screening},
it is important to take into account screening for a correct description of the
density and temperature dependence of the low-temperature conductivities in 2D
systems.

\subsubsection{Dielectric function}
\label{sec:dielectric}

The long-wavelength dielectric function of a doped 2D semiconductor has two
contributions: the first is due to the gate-induced \emph{extrinsic} carriers
which is normally accounted for in transport studies, and the second from the
\emph{intrinsic} screening associated with virtual electron-hole (eh) pair
excitation between the valence and conduction bands. Contrary to 3D bulk systems
where the latter is well described by the dielectric constant of the material,
intrinsic screening in 2D semiconductors is strongly $q$
dependent~\cite{Rubio:Dielectric,Louie:Screening}.

The total dielectric function thus becomes
\begin{equation}
  \label{eq:eps_tot}
  \epsilon(q,T,\mu) = 1 - V(q) \chi_\text{ext}(q,T,\mu) 
                        - V(q) \chi_\text{int}(q)
\end{equation}
where $V(q)=e^2/2\epsilon_0q$ is the 2D Fourier transform of the Coulomb
interaction and $\chi_\text{ext}$ ($\chi_\text{int}$) is the polarizability due
to intraband (interband) eh-pair processes involving \emph{extrinsic}
(\emph{intrinsic}) carriers. Dielectric screening from the environment is
included with the replacements $\epsilon\rightarrow \kappa\epsilon$ and
$V\rightarrow V/\kappa$, where $\kappa=4.1$ is an effective background
dielectric constant for encapsulation in $h$-BN~\cite{Jena:ChargeScattering}.

% \subsubsection{Extrinsic carrier screening}

The screening from the \emph{extrinsic} carriers in the valence/conduction band
is described with finite-temperature RPA theory where the static polarizability
at finite temperatures is given by~\cite{Maldague:ManyBody,Stern:2D}
\begin{equation}
  \label{eq:maldague}
  \chi_\text{ext}(q, T,\mu) = \int_0^\infty \! d\mu' \, 
      \frac{\chi_\text{ext}(q, 0, \mu')}{4k_\text{B}T \cosh^2{\frac{\mu -
            \mu'}{2k_\text{B}T}}}  .
\end{equation}
Here, $\chi_\text{ext}(q, 0, \mu)$ is the zero-temperature polarizability and
the integral is evaluated following Ref.~\onlinecite{Flensberg:Plasmon}. For a
degenerate 2DEG, Eq.~\eqref{eq:maldague} yields the above-mentioned TF result
for the extrinsic screening. In the nondegenerate regime, extrinsic carrier
screening is described by Debye-H{\"u}ckel theory and is strongly reduced. Thus,
the strength of the Coulomb potential increases dramatically, and the transport
is expected to become dominated by Coulomb-disorder scattering at low carrier
density and high temperature.

% \subsubsection{Intrinsic screening}

For the intrinsic screening due to interband eh-pair excitations between the
valence and conduction bands, we use the dielectric
function~\cite{Rubio:Dielectric,Louie:Screening}
\begin{align}
  \label{eq:epsilon_2D}
  \epsilon_\text{int}(q) & = 1 - \frac{e^2}{2\epsilon_0 q}  \chi_\text{int}(q) 
  \nonumber \\
                      & = 1 + \alpha q ,
\end{align}
which is valid in the long-wavelength limit and where $\alpha$ is given by the
interband polarizability $\chi_\text{int}(q)\propto q^2$, and
$\alpha \approx 40$~{\AA} for
MoS$_2$~\cite{Thygesen:Excitons,Thygesen:Simple}. Note that the intrinsic
screening vanishes, $\epsilon_\text{int} \rightarrow 1$, for $q\rightarrow 0$ in
2D semiconductors. As it originates from eh-pair excitations between bands
separated by a large band gap, the intrinsic screening is not expected to depend
on temperature.

\subsubsection{Extrinsic vs intrinsic screening}

The dielectric function in Eq.~\eqref{eq:eps_tot} is similar to the one in gated
graphene~\cite{Sarma:Dielectric}. However, due the sizeable band gaps in 2D
TMDs, \emph{intrinsic} screening is much weaker than in graphene, and the
long-wavelength dielectric function is dominated by intraband processes
involving \emph{extrinsic} carriers.

Already from the $q$ dependence of the polarizabilities it is clear that
extrinsic screening dominates the total dielectric function in
Eq.~\eqref{eq:eps_tot} in the long-wavelength limit. For a degenerate carrier
distribution, the wave vector $q^*$ at which the two screening mechanisms
contribute equally to the total dielectric function can be estimated by
comparing the products of the Coulomb interaction and the polarizabilities,
\begin{equation}
  \label{eq:screening}
  q_\text{TF}/q^* = \alpha q^*  \quad \rightarrow \quad
  q^* = \sqrt{q_\text{TF} / \alpha} .
\end{equation}
For monolayer MoS$_2$, we estimate $q^* \approx 0.3\,\pi/a$ based on the
conduction-band TF wave vector. Thus, extrinsic screening dominates at
temperatures where the scattering wave vector is limited by $q<2k_F\ll q^*$,
whereas at (i) high temperatures $T\gg T_F$ where extrinsic carrier screening is
reduced, and (ii) high carrier densities where $k_F$ becomes appreciable
compared to $q^*$, intrinsic screening becomes important.

\section{Results}

% \begin{figure*}[!t]
%   \centering
%   \includegraphics[width=1.0\linewidth]{momobilities} 
%   \caption{Hole and electron mobilities limited by \emph{charged} Mo
%     vacancies. The plots show the hole (left) and electron (right) mobilities vs
%     carrier density and temperature for a disorder density of
%     $n_\text{Mo}=10^{11}$~cm$^{-2}$. The corresponding density,
%     $\alpha = d\mathrm{log}(\mu)/d\mathrm{log}n$, and temperature,
%     $\gamma = -d\mathrm{log}(\mu)/d\mathrm{log}T$, exponents of the mobilities
%     are shown in the lower row. The weak dashed lines show the mobility in the
%     absence of intrinsic screening. The dots ($\bullet$) indicate the
%     temperatures and densities where $T = T_F$.}
% \label{fig:mu_Mo}
% \end{figure*}
In the following, we present our results for the density, $\mu\sim n^\alpha$,
and temperature, $\mu\sim T^{-\gamma}$, dependencies of the mobility limited by
atomic vacancies. As the inverse relaxation time scales directly with the
disorder density, we restrict here the discussion to
$n_\text{dis} = 10^{11}$~cm$^{-2}$ which corresponds to a dilute
concentration ($c_\text{dis} \approx 0.01\,\%$) of defects. 

From our considerations in the previous section, we anticipate a density scaling
of the mobility with (i) $\alpha<0$ if the short-range potential dominates
as the scattering rate increases with the electron and hole energies
[cf. Fig.~\ref{fig:Tmatrix} and Eq.~\eqref{eq:tau_tmatrix}], and (ii)
$\alpha \gtrsim 0 $ if the Coulomb potential
dominates~\cite{Hwang:Short,Hwang:Universal}.
\begin{figure*}[!t]
  \centering
  \includegraphics[width=1.0\linewidth]{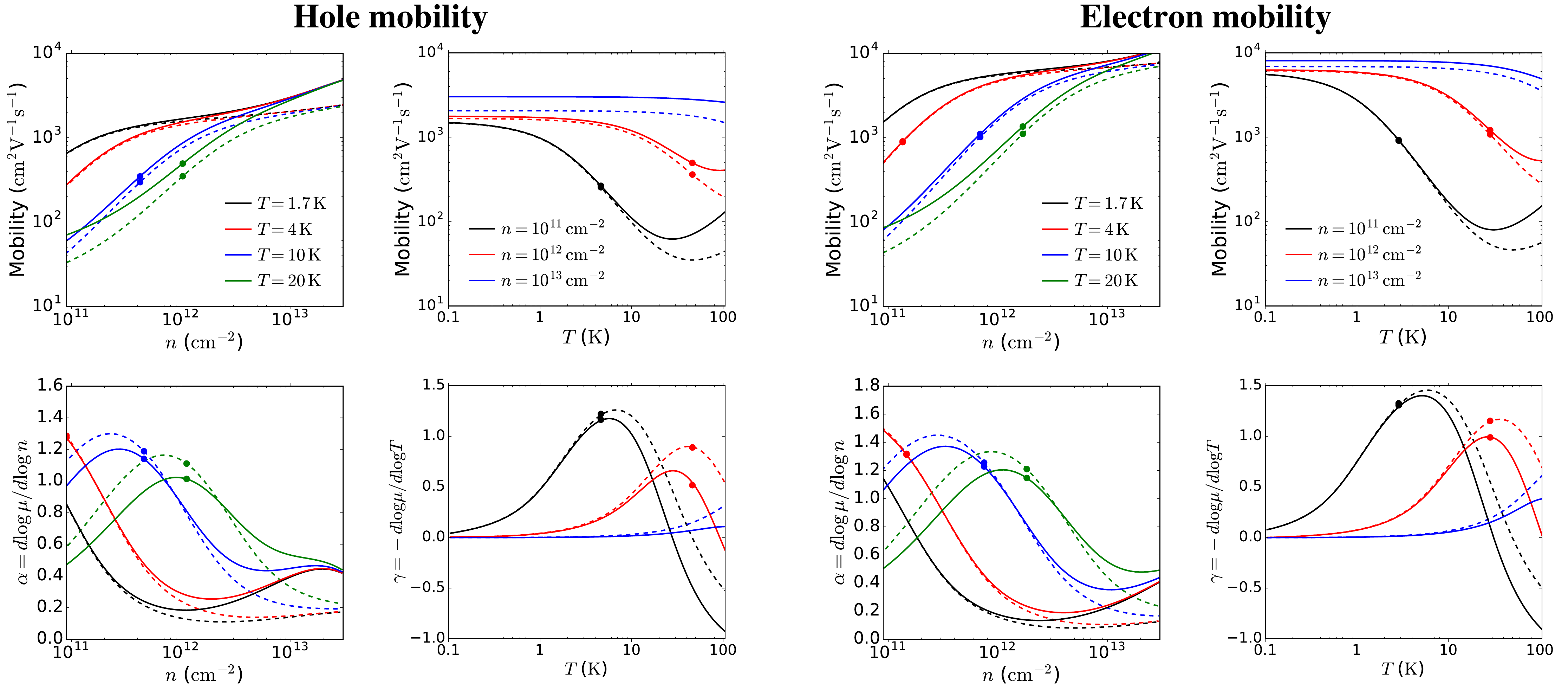} 
  \caption{Hole and electron mobilities limited by \emph{charged} S
    vacancies. The plots show the hole (left) and electron (right) mobilities
    vs carrier density and temperature for a disorder density of
    $n_\text{S}=10^{11}$~cm$^{-2}$. The corresponding density,
    $\alpha = d\mathrm{log}\mu/d\mathrm{log}n$, and temperature,
    $\gamma = -d\mathrm{log}\mu/d\mathrm{log}T$, exponents of the mobilities are
    shown in the lower row. The weak dashed lines show the mobility in the
    absence of intrinsic screening. The dots ($\bullet$) indicate the
    temperatures and densities where $T = T_F$.}
\label{fig:mu_S}
\end{figure*}

\subsection{Charged vacancies: Coulomb-disorder dominated transport}
\label{sec:coulomb}

In Fig.~\ref{fig:mu_S} we show the density and temperature dependence of the
mobility limited by \emph{charged} vacancies. As we explain below, the results
for charged Mo and S vacancies are almost identical, hence only the results for
S vacancies are shown. The scaling exponents
$\alpha = d\mathrm{log}\mu/d\mathrm{log}n$ and
$\gamma=-d\mathrm{log}\mu/d\mathrm{log}T$ are shown in the bottom panels. To
emphasize effect of the intrinsic interband screening, we show the results both
with (full lines) and without (dashed lines; $\epsilon_\text{int}=1$) intrinsic
screening, and the dots indicate the densities and temperatures where the Fermi
temperature is equal to the actual temperature, $T = T_F$, and hence mark the
quantum-classical crossover between the degenerate ($T< T_F$) and nondegenerate
($T > T_F$) regimes.

Overall, the mobilities limited by charged S vacancies show a strong density and
temperature dependence with $\alpha >0$ and $\gamma > 0$, except at high
temperatures and low densities where $\gamma<0$. These observations indicate
that the vacancy-independent Coulomb potential scattering dominates the inverse
relaxation time for charged vacancies, thus resulting in almost identical
mobilities for charged Mo and S vacancies. Based on the observations regarding
the energy dependence of the relaxation time in Eq.~\ref{eq:tau_tmatrix}, a
$\alpha < 0$ density scaling would have been expected if the short-range
potential was dominating.

This is also consistent with the observed strong temperature dependence of the
mobility which stems from the temperature-dependent screening of the Coulomb
potential. Moving from the quantum-classical crossover into the nondegenerate
low-density/high-temperature regimes, the strength of the Coulomb potential
increases dramatically due to a strong reduction in the efficiency of extrinsic
carrier screening. In turn, this leads to a pronounced reduction of the mobility
of up to two orders of magnitude.

The effect of intrinsic screening on the Coulomb potential is evident at high
temperatures $T>T_F$ where it becomes comparable to extrinsic screening. Here,
it gives rise to an unusual upturn in the mobility with increasing temperature
(compare full and dashed lines). Also, at high densities the mobilities show a
pronounced upturn with increasing density. This is again due to the intrinsic
screening which becomes stronger with increasing density ($k_F$) due to its
linear $q$ dependence in~\eqref{eq:epsilon_2D}. Overall, the differences between
the full and dashed lines show that intrinsic screening gives rise to a
significant boost of the mobility in the nondegenerate regime and for high
densities in the degenerate regime.

Finally, we note that the hole mobilities are in general lower than the electron
mobilities in the degenerate regime. This can be attributed to the lack of spin
degeneracy in the valence band which results in a weaker screening of Coulomb
scattering for holes (cf. Sec.~\ref{sec:VC}).

\begin{figure}[!t]
  \centering
  \includegraphics[width=0.65\linewidth]{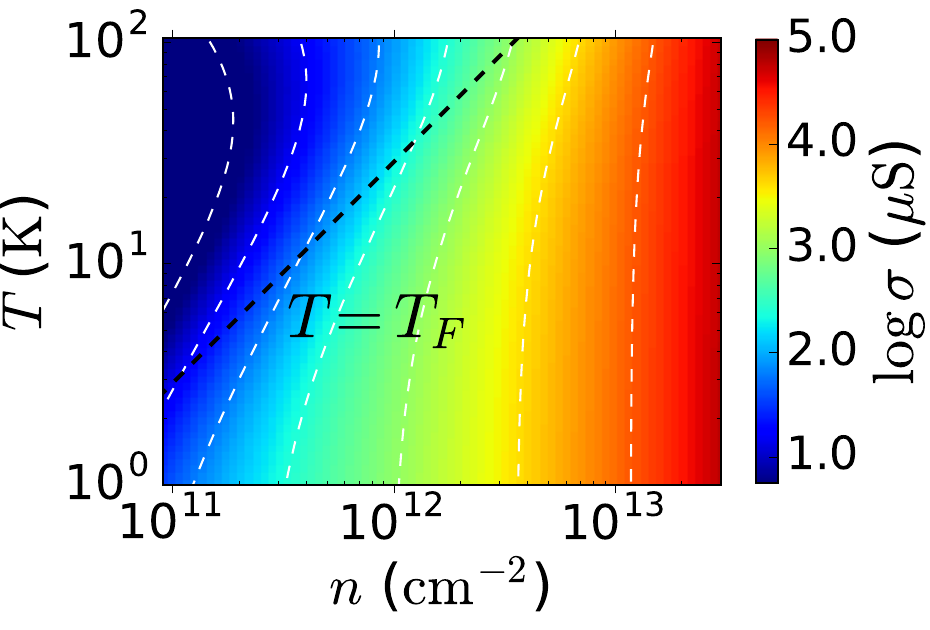}
  \caption{Transport diagram for $n$-type 2D MoS$_2$ showing the conductivity
    $\sigma$ limited by \emph{charged} S vacancies
    ($n_\text{S}=10^{11}$~cm$^{-2}$) as a function of carrier density $n$ and
    temperature $T$.}
\label{fig:phase}
\end{figure}

The density and temperature range where the pronounced screening-induced change
in the mobilities in Fig.~\ref{fig:mu_S} takes place, is summarized in the
transport diagram shown for the S-vacancy limited conductivity in $n$-type
MoS$_2$ in Fig.~\ref{fig:phase}. The diagram shows the conductivity
$\sigma= e^2 n \expect{\tau}/m^*$ (here $\log\sigma$) in the $(n,T)$ plane and
the black dashed line indicates the quantum-classical crossover at $T=T_F$. As
is evident, the crossover takes place at experimentally easily accessible
densities and temperatures.

The quantum-classical crossover is clearly visible in the conductivity data. In
the degenerate high-density regime ($T \ll T_F$), the conductivity scales as
$\sigma \sim n$ and is almost independent on temperature due to the effective
short-range nature of the Coulomb potential in Eq.~\eqref{eq:VC0} caused by
carrier screening. At $T\gtrsim T_F$, the conductivity acquires a stronger
density and temperature dependence due to the reduced screening of the Coulomb
potential.

\subsection{Neutral vacancies: Short-range disorder limited transport}
\label{sec:nocoulomb}

In this section, we consider the situation where the vacancies remain neutral in
doped samples. This is expected to occur in 2D TMDs where vacancies do
not introduce occupied in-gap states (e.g., $X=\text{S, Se}$ vacancies in
W$X_2$~\cite{Sanyal:Systematic,Huis:Strong}), implying that the vacancy will
remain overall neutral in $p$-doped samples and only the short-range
contribution to the relaxation time in Eq.~\eqref{eq:tau_total} remains. 
% This is unlikely to take place in $n$-doped samples, as vacancies always
% introduce empty midgap states.

\begin{figure}[!t]
  \centering
  \includegraphics[width=0.65\linewidth]{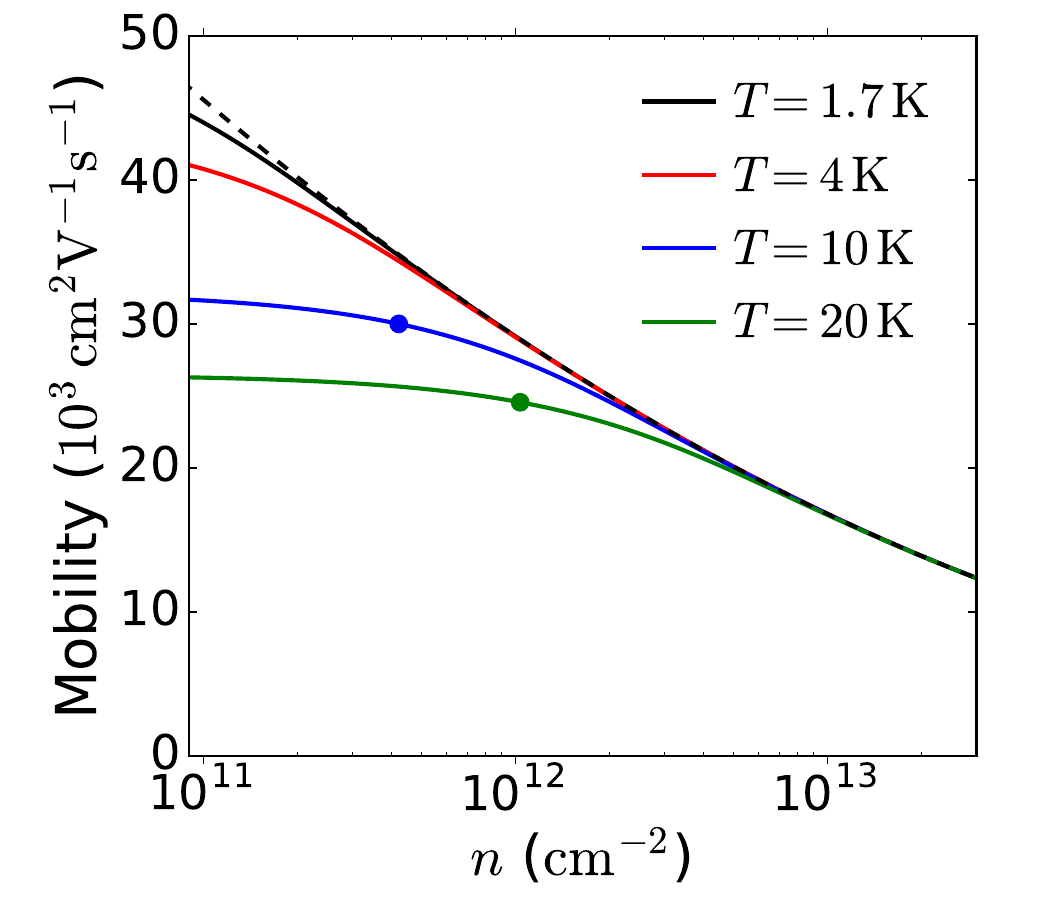}
  \caption{Hole mobility in 2D MoS$_2$ limited by \emph{neutral} S vacancies,
    $n_\text{S}=10^{11}\,$cm$^{-2}$, at different temperatures. The dashed line
    shows the zero-temperature mobility
    $\mu_{xx}=e \tau_{i,\sigma}(E_F) / m_\sigma^*$ with $\tau_{i,\sigma}$ given
    by its $T$-matrix form in Eq.~\eqref{eq:tau_tmatrix}.}
\label{fig:mobility_neutral}
\end{figure}

To demonstrate the impact on the hole mobility in TMDs where the above-mentioned
situation is realized, we show in Fig.~\ref{fig:mobility_neutral} the hole
mobility in MoS$_2$ limited by \emph{neutral} S vacancies at different
temperatures. The dashed line shows the zero-temperature mobility
$\mu=e \tau_{i,\sigma}(E_F) / m_\sigma^*$ with the relaxation time obtained from
Eq.~\eqref{eq:tau_tmatrix}. As the semiconducting 2D TMDs have similar valence
band structure, only minor quantitative changes are expected for hole mobility
limited by \emph{neutral} $M$ and $X$ vacancies in other TMDs.

First of all, we note that the mobilities are more than an order of magnitude
higher than the ones in Fig.~\ref{fig:mu_S} limited by charged vacancies, except
at the highest carrier densities where they become comparable. The fact that the
mobilities are higher for neutral vacancies, is not surprising because the
Coulomb potential completely dominates the short-range potential for scattering
off charged vacancies. As we had anticipated from the scattering rates deduced
from Fig.~\ref{fig:Tmatrix}, the mobilities in Fig.~\ref{fig:mobility_neutral}
limited alone by the short-range potential due to neutral vacancies, decrease
with increasing carrier density. This rather unusual density dependence of the
mobility is a direct consequence of the $T$-matrix induced renormalization of
the scattering amplitude. At low densities, the decreasing temperature
dependence of the mobility stems from the energy average of the relaxation time
in Eq.~\eqref{eq:tau_averaged}, which with increasing temperature probes the
increasing energy dependence of the scattering rate at higher energies. The
resulting reduction in the mobility is, however, much smaller than the
sceening-related reduction observed for charged vacancies.

\begin{figure}[!t]
  \centering
  \includegraphics[width=0.65\linewidth]{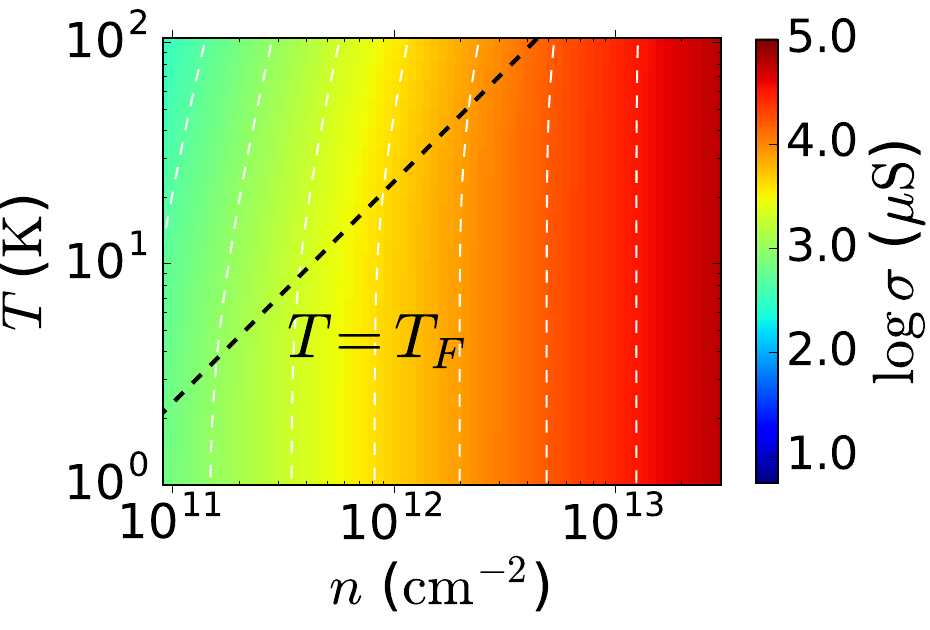}
  \caption{Transport diagram for $p$-type 2D MoS$_2$ showing the conductivity
    $\sigma$ limited by \emph{neutral} neutral S vacancies
    ($n_\text{S}=10^{11}$~cm$^{-2}$) as a function of carrier density $n$ and
    temperature $T$.}
\label{fig:phase_p}
\end{figure}

The qualitative difference between the screening-induced and $T$-matrix induced
density and temperature dependencies of the mobility emerges clearly from the
transport diagrams for charged and neutral vacancies in Figs.~\ref{fig:phase}
and~\ref{fig:phase_p}, respectively. While in the latter case, the conductivity
$\sigma \sim n^\alpha$ scales roughly as $\alpha\approx 1$ on both sides of the
quantum-classical crossover, the screening-induced density dependence of the
scattering rate in the former case results in a stronger $\alpha\approx 2$
behavior in the nondegenerate regime at $T\gtrsim T_F$.

These qualitative differences in the transport characteristics allow to clearly
distinguish Coulomb-disorder limited transport from short-range limited
transport due to, e.g., charged and neutral vacancies, respectively.

\section{Discussion}

Reported experimental low-temperature mobilities in 2D TMDs have thus far been
rather
low~\cite{Herrero:Intrinsic,Kis:Engineering,Eda:Transport,Wang:Towards,Hone:Multi,Wang:High,Wang:Realization},
and have only recently exceeded
$\sim$1000~cm$^{-2}\,$V$^{-1}\,$s$^{-1}$~\cite{Tutuc:Shubnikov,Ensslin:Gate,Dean:Ambi,Tutuc:Large,Ensslin:Interactions}.
This is still slightly lower than our calculated mobilities limited by charged
vacancies and much lower than the calculated mobilities limited by neutral
vacancies. This indicates that the concentration of vacancies in experimental
samples could be higher than $n_\text{dis} = 10^{11}$~cm$^{-2}$ used
here. Several reports of high concentrations of S
vacancies~\cite{Wang:Hopping,Wang:Towards} suggest that this is indeed a likely
reason for the low experimental mobilities. Only recently have defect densities
of $n_\text{dis} \sim 10^{11}$~cm$^{-2}$ been
demonstrated~\cite{Pasupathy:Approaching}. However, experimental transport
properties of such high-quality TMDs have so far not been reported.

The often observed strong metallic temperature and density dependence of
experimental low-temperature
mobilities~\cite{Herrero:Intrinsic,Kis:Engineering,Eda:Transport,Wang:Towards,Hone:Multi,Wang:High,Wang:Realization}
resemble best our results for charged vacancies. This indicates that a
combination short-range and Coulomb-disorder scattering is limiting the
low-temperature mobility in 2D TMDs. As we here suggest, vacancies may be the
source of both short-range and Coulomb disorder scattering. Hence, there is no
need to introduce additional Coulomb
disorder~\cite{Fischetti:Mobility,Jena:ChargeScattering} (e.g., charged
impurities in the substrate) in order to account for the metallic transport
behavior.

\section{Conclusions}

In conclusion, we have studied the effect of atomic vacancies on carrier
scattering and transport in $p$- and $n$-type monolayer MoS$_2$. Due to the
presence of both filled and empty vacancy-induced in-gap states, the vacancies can
be expected to become \emph{charged} in $p$- and $n$-doped MoS$_2$, and thereby
give rise to both short-range and Coulomb disorder scattering. The situation is
similar for vacancies in many other 2D TMDs, but cases lacking filled in-gap
states have been reported, implying that the vacancy remain \emph{neutral} in
the $p$-doped material.

Studying the short-range scattering properties of vacancies with the
$T$-matrix formalism, we show that multiple scattering gives rise to a strong
renormalization of the Born scattering amplitude, which results in a pronounced
reduction as well as a characteristic energy dependence of the scattering rate.
As a result, the Coulomb contribution to the scattering potential for
\emph{charged} vacancies by far dominates carrier scattering. This results in a
strong screening-induced temperature and density dependence of the mobility in
2D TMDs hosting charged vacancies. For TMDs with \emph{neutral} vacancies the
mobility is significantly higher and shows an unusual behavior with a decreasing
density dependence.
% reaches values $>10.000$~cm$^{-2}\,$V$^{-1}\,$s$^{-1}$ for a vacancy
% concentration of $n_\text{dis}= 10^{11}$~cm$^2$ and low temperatures
Thus, TMDs in which vacancies remain neutral in the doped material are better
candidates for high-mobility devices.

\begin{acknowledgments}
  K.K. acknowledges support from the Carlsberg Foundation and the European
  Union's Horizon 2020 research and innovation program under the Marie
  Sklodowska-Curie Grant Agreement No.~713683 (COFUNDfellowsDTU). T.L.
  acknowledges partial support from NSF ECCS-1542202. The Center for
  Nanostructured Graphene (CNG) is sponsored by the Danish National Research
  Foundation, Project DNRF103.
\end{acknowledgments}

% \bibliography{journalabbreviations,references}
'\bibliography{paper}

\begin{thebibliography}{95}
\bibitem [{cal()}]{calculations}%
  \BibitemOpen
  \href@noop {} {}\bibinfo {note} {The DFT calculations of the band structure,
  effective masses, and defect matrix elements have been performed with the
  electronic structure code GPAW~\cite{GPAW,GPAW1,GPAW2} within the projector
  augmented-wave method, using the LDA xc functional and a LCAO double-zeta
  polarized (DZP) basis set. The defect potentials were obtained using a $11
  \times 11$ supercell with 10~{\AA} of vacuum between the MoS$_2$ sheet and
  the cell boundaries in the vertical direction imposed with Dirichlet boundary
  conditions in order to avoid spurious interlayer interactions. Lastly, the
  defect matrix elements and $T$ matrix were calculated on a $135\times 135$ BZ
  grid using a numerical broadening of $\eta=10$~meV.}\BibitemShut {Stop}%
\end{thebibliography}

\end{document}